\documentclass[a4paper]{article}
\usepackage{INTERSPEECH2022}
\usepackage{amsmath,graphicx}
 \usepackage{booktabs}
 \usepackage{array}
 \usepackage[bookmarks=false]{hyperref}
 \usepackage{xcolor}
 \usepackage{cleveref}
\usepackage{cite}

\title{Mixed-Phoneme BERT: Improving BERT with Mixed Phoneme and Sup-Phoneme Representations for Text to Speech}

\makeatletter
\def\@name{ \emph{Guangyan Zhang$^1$, Kaitao Song$^2$, Xu Tan$^{2,*}$\thanks{$^*$ Work done during the first author is interning at Microsoft. Corresponding author: Xu Tan, xuta@microsoft.com.}, Daxin Tan$^1$, Yuzi Yan$^3$,} \\ \emph{ Yanqing Liu$^4$, Gang Wang$^4$,Wei Zhou$^5$, Tao Qin$^2$, Tan Lee$^1$, Sheng Zhao$^4$}}
\makeatother

\address{
  $^1$ Department of Electronic Engineering, The Chinese University of Hong Kong\\
  $^2$ Microsoft Research Asia $^3$ Tsinghua University $^4$ Microsoft Azure Speech $^5$ Zhejiang University 
}

\email{gyzhang@link.cuhk.edu.hk, \{kaitaosong, xuta\}@microsoft.com}

\begin{document}
%\ninept
\maketitle
\ninept
\begin{abstract}
Recently, leveraging BERT pre-training to improve the phoneme encoder in text to speech (TTS) has drawn increasing attention. However, the works apply pre-training with character-based units to enhance the TTS phoneme encoder, which is inconsistent with the TTS fine-tuning that takes phonemes as input. Pre-training only with phonemes as input can alleviate the input mismatch but lack the ability to model rich representations and semantic information due to limited phoneme vocabulary. In this paper, we propose Mixed-Phoneme BERT, a novel variant of the BERT model that uses mixed phoneme and sup-phoneme representations to enhance the learning capability. Specifically, we merge the adjacent phonemes into sup-phonemes and combine the phoneme sequence and the merged sup-phoneme sequence as the model input, which can enhance the model capacity to learn rich contextual representations. Experiment results demonstrate that our proposed Mixed-Phoneme BERT significantly improves the TTS performance with 0.30 CMOS gain compared with the FastSpeech 2 baseline. The Mixed-Phoneme BERT achieves $3\times$ inference speedup and similar voice quality to the previous TTS pre-trained model PnG BERT.
%The Mixed-Phoneme BERT has been deployed on Microsoft Azure TTS services.%

\end{abstract}
\noindent\textbf{Index Terms}: Text to Speech,  Pre-training, BERT, Sup-Phoneme, Mixed representation

\vspace{-0.7em}

\section{Introduction} 
\label{sec:intro}

In recent years, neural text to speech (TTS) ~\cite{tan2021survey,tan2022naturalspeech,wang2017tacotron,shen2018natural,ren2019fastspeech,ren2020fastspeech, zhang2021study} has demonstrated significant successes in producing natural-sounding speech. Specifically, the non-autoregressive TTS systems\cite{ren2019fastspeech} have received increasing attention due to their advanced ability in generating stable mel-spectrograms with fast speed. However, existing non-autoregressive TTS still retains some flaws, like the ``one-to-many'' mapping between text and speech. In \cite{ren2020fastspeech, elias21_interspeech, zhang21u_interspeech}, the information from the ground truth speech data, e.g. pitch, duration, is incorporated in model training to alleviate the ``one-to-many'' problem. It was noted that the generated speech still tends to carry flat prosody since the phoneme sequence does not contain adequate information for predicting natural prosody \cite{kenter2020improving}.

\vspace{-0.3em}

In order to handle this problem, some works have tried to enrich the phoneme sequence input with syntactic information and linguistic features~\cite{taylor2009text, zen2015unidirectional,wu2016merlin}. However, this strategy has two main disadvantages: i) it needs several separated text pre-processing modules (e.g., POS tagger and syntax parser), possibly leading to error propagation from one module to the subsequent ones; ii) designing and labelling linguistic features are usually time-consuming tasks that require language expertise~\cite{zhang2020learning}. Recently, leveraging contextual representations learned from unlabeled text data to improve the TTS model has become a rising topic \cite{wang2015word,chung2019semi,hayashi2019pre,kenter2020improving,xu2021improving}, which is also the focus of our paper.
\vspace{-0.3em}

Pre-trained language models (e.g., BERT~\cite{devlin2019bert}) have achieved state-of-the-art performance in solving natural language processing tasks. Recent works~\cite{hayashi2019pre,xiao2020improving,kenter2020improving,xu2021improving} have applied the BERT model as an auxiliary encoder for the TTS system. The auxiliary BERT encoder extracts additional text features for character-based units (e.g., character\cite{el2020characterbert}, subword\cite{liu2019roberta, devlin2019bert}), which enables the TTS system to generate speech with better pronunciation and expressiveness.  Those methods have to face the challenge that the pre-training with the character-based units is inconsistent with the TTS fine-tuning taking pure phoneme as the input. The inconsistency will bring several problems: 1)  The alignment between phoneme and character information might be unstable(e.g., misalignment between phoneme and character sequence), making the TTS phoneme encoder fail to leverage the pre-trained text features. 2) The auxiliary BERT encoder with text input will also bring longer training/inference time and more computing resources. Besides, using phoneme and the auxiliary encoder simultaneously will result in larger model parameters. Some works \cite{kastner2019representation, jia21_interspeech} attempt to enhance the TTS phoneme encoder with character information directly rather than introducing an auxiliary BERT model. Since the model still needs character-based units as an extra input, inconsistency and corresponding drawbacks still exist.

% \vspace{0.3em}

So what if we pre-train the phoneme encoder with only the phoneme as the input? We note that the size of the phoneme dictionary is only around 200, and directly using such a small dictionary would not convey contextual semantic information effectively\cite{ding2019call}. To enhance the representation capacity of model input and avoid the problem of inconsistency between pre-training and fine-tuning, we propose Mixed-Phoneme BERT, a novel variant of the BERT model to handle TTS tasks. The Mixed-Phoneme BERT model merges phoneme and sup-phoneme sequences into a new sequence. The sup-phoneme tokens are obtained by applying the learnt Byte-Pair Encoding (BPE) ~\cite{sennrich2016neural} rules to words. Compared with the phoneme dictionary, which is usually too small, the dictionary size of sup-phoneme is much larger, producing representations with better semantics. The mixed-Phoneme BERT employs Masked Language Modelling objective for pre-training on large-scale unlabeled text corpora. In order to prevent the information leakage problem, a pre-processed data alignment and consistent masking strategy are introduced, which require our Mixed-Phoneme BERT model to predict the masked sup-phoneme tokens and all phoneme tokens corresponding to the masked sup-phoneme tokens simultaneously. The pre-trained Mixed-Phoneme BERT serves as the phoneme encoder for TTS fine-tuning. Experiment results indicate that the Mixed-Phoneme BERT model can significantly improve the TTS performance over the baseline Fastspeech 2 model and alleviate the flat prosody problem. Ablation studies also validate the effectiveness of using the sup-phoneme information in the input representations.

 \begin{figure*}[t]
    \captionsetup{font=scriptsize}
  \centering
  \includegraphics[width=\textwidth]{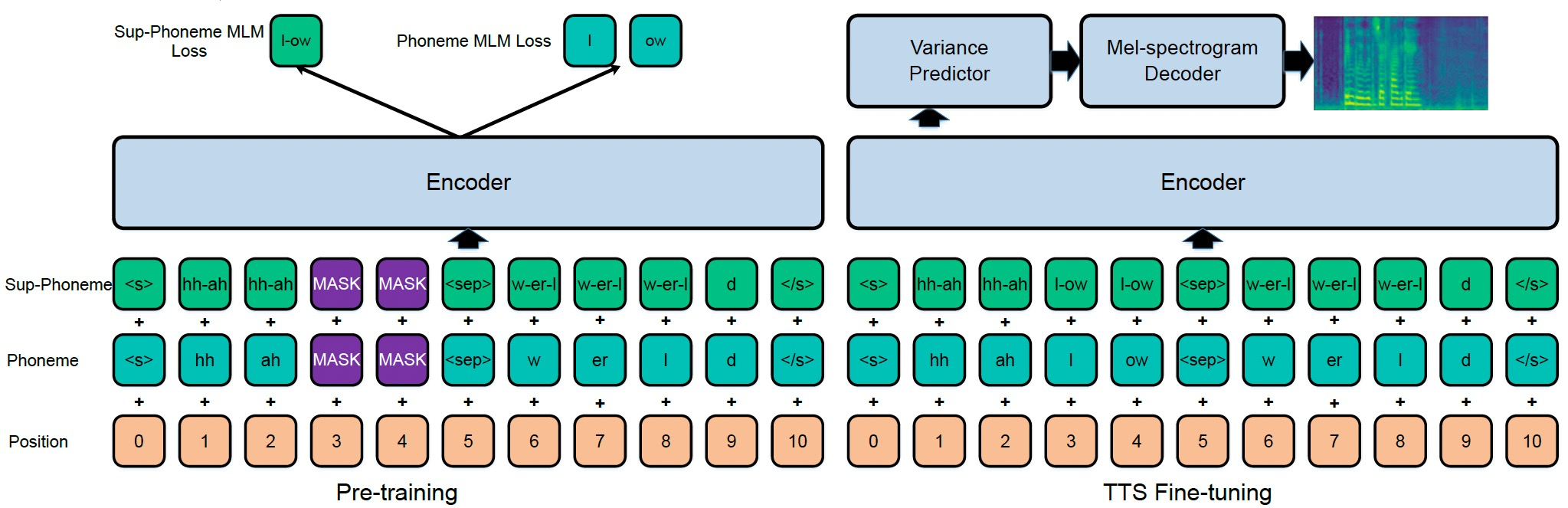}
  \caption{Pre-training and TTS fine-tuning of Mixed-Phoneme BERT. The position, phoneme, and sup-phoneme embeddings are illustrated in different colors. A token in purple means that masking is applied.  Two example word segments, i.e., ``hello" and ``world", are shown. ``hello" consists of two sup-phoneme units, \textbf{hh-ah} and \textbf{l-ow}. The sup-phoneme unit \textbf{l-ow} is masked during pre-training. }
  \label{Figure:model_arch}
\end{figure*}

\vspace{-0.3em}

\section{Mixed-Phoneme BERT}

\subsection{Overview}
The proposed Mixed-Phoneme BERT model is specifically designed for the TTS tasks, which is composed of two stages: pre-training and TTS fine-tuning, just as shown in \autoref{Figure:model_arch}. We first pre-train Mixed-Phoneme BERT on large amounts of unlabeled text data, then employ paired speech-text for TTS fine-tuning. Specifically, the input representation to the Mixed-Phoneme BERT comprises two different types of tokens derived from text, namely phonemes and sup-phonemes. In pre-training, the sup-phoneme tokens and corresponding phoneme tokens are masked randomly to make predictions. At the TTS fine-tuning stage, the token sequences without any masking are used as input, and the pre-trained BERT model serves as the phoneme encoder of TTS. More details are given in the following sections.

%\vspace{0.3em}

\subsection{Mixed Phoneme and Sup-Phoneme Representations}
Most existing TTS systems use the phoneme sequence to represent the input text. Only phoneme inputs can only provide pronunciation while lacking the ability to capture contextual representations. Moreover, directly using words or other similar types as the TTS model inputs will lead to large size of token inventory, which is undesirable for model training, and the derived out-of-vocabulary problem may also hinder the system's operation. Therefore, a mixed representation of phoneme and sup-phoneme sequences is proposed for pre-training of the Mixed-Phoneme BERT. A sup-phoneme refers to a group of neighbouring phonemes and does not necessarily correspond or relate to a lexical word. Inspired by the success of byte-pair encoding (BPE) in handling the out-of-vocabulary problem\cite{sennrich2016neural}, we apply BPE to encode each word into one or more sup-phoneme token(s) and thus obtain the final sup-phoneme.

%A mixed representation of phoneme and sup-phoneme sequences is proposed for pre-training of the Mixed-Phoneme BERT. A sup-phoneme refers to a group of neighboring phonemes, and does not necessarily correspond or relate to a lexical word. 

%Most existing TTS systems use phoneme sequence to represent the input text. However, individual phonemes do not capture contextual representations. Using words or like-size units as TTS model input would lead to a very large size of token inventory, which is undesirable for model training. The out-of-vocabulary problem may also hinder the system's operation. Inspired by the success of byte-pair encoding (BPE) in handling the out-of-vocabulary problem\cite{sennrich2016neural}, we apply BPE to encode each word into one or more sup-phoneme token(s). 

Byte-Pair Encoding (BPE) is first introduced for word segmentation in \cite{sennrich2016neural}. In \cite{sennrich2016neural} , each word is represented as a sequence of characters. While in this paper, each word can be seen as a sequence of phonemes, and phoneme is the basic unit for BPE learning. The BPE algorithm consists of learning and encoding stages. During BPE learning\cite{wolf2020transformers}, each word in the training corpus is converted to a sequence of phonemes. A base vocabulary that consists of BPE merge rules is learnt iteratively. This learning process repeats until the base vocabulary reaches the pre-defined size. The learned BPE rules can be applied to encode a word into the sup-phoneme token(s) at the encoding process. The encoding process is greedy (utilizing large sup-phoneme tokens as much as possible) and deterministic.

Due to our design, the sup-phoneme has a larger dictionary size than the phoneme, thus supporting higher representation capacity in representing word semantics. In addition, we can easily align each sup-phoneme token to its corresponding phoneme tokens, which is beneficial for us to mix phoneme and sup-phoneme representations. More in detail, when given the text input, considering the sup-phoneme sequence are usually shorter than the phoneme sequence, we up-sample each sup-phoneme token based on the number of its corresponding phoneme tokens. The overall mixed phoneme and sup-phoneme representations for each input sentence are formed by summing up the corresponding up-sampled sup-phoneme sequence, phoneme sequence and position embeddings. The phoneme sequence is used to enhance the pronunciation for the TTS task, while the sup-phoneme sequence can bring semantic and contextual information for the model.

\subsection{Pre-training}
 \label{sec:format}
The pre-training procedure for Mixed-Phoneme BERT is similar to that for the original BERT, which applies self-supervised learning on unlabeled text for training. During pre-training, a certain percentage of input tokens are randomly selected and replaced with the special token {\ttfamily MASK}, and then the masked tokens are predicted by the Masked Language Modelling (MLM) objective.

As aforementioned, the phoneme and sup-phoneme sequences are used to represent the same text content. If random masking is applied to the two sequences independently, an information leakage problem may occur such that the model tends to infer the masked tokens in one sequence through the corresponding unmasked tokens in the other sequence. As a result, the model cannot effectively capture contextual information, and the learned representations will be useless. To avoid this problem, we apply a consistent masking strategy to mask sup-phoneme and phoneme sequence, which is if a sup-phoneme token is chosen to be masked, its corresponding phoneme tokens will also be masked. Following previous works~\cite{liu2019roberta}, $15\%$ of the sup-phoneme tokens (except special tokens) in each sequence are masked at random. If a sup-phoneme token is chosen to be masked, it is given: i) $80\%$ chance of keeping the {\ttfamily MASK} symbol; ii) $10\%$ chance to be replaced by a random sup-phoneme token; iii) $10\%$ chance of remaining unchanged. 

% \vspace{0.3em}
Additionally, two masking strategies are evaluated, i.e., with or without whole word masking~\cite{chinese-bert-wwm}. When the whole word masking is used, if a masked sup-phoneme token belongs to a word, all sup-phoneme tokens that make up that word are masked together. This technique compels the model explicitly to recover the whole word, making the MLM pre-training task more challenging.

% \vspace{0.3em}
The MLM objective is used to predict the masked tokens with cross-entropy loss. The final hidden vectors of the encoder corresponding to the masked phoneme tokens are selected for the masked phoneme tokens classification. Then the hidden vectors of the masked phoneme tokens are aggregated to the sup-phoneme level by average pooling for the masked sup-phoneme tokens classification. The sup-phoneme and phoneme level cross-entropy losses are added together as the final objective.

\vspace{-1em}

\section{Experimental Setup}

\subsection{Datasets}
\subsubsection{Pre-training Data}
Pre-training data is used to train the Mixed-Phoneme BERT and learn BPE base vocabulary. The pre-training data is composed of WMT news and Wikipedia corpora, including 83 million sentences from the WMT news and Wikipedia corpora. The dataset is divided into 81M sentences for training, 1M for validation, and 1M for testing. 

\subsubsection{TTS Fine-tuning Data}
 We choose the LJSpeech corpus as the benchmark to evaluate the pre-trained Mixed-Phoneme BERT model during the TTS fine-tuning stage. The LJSpeech contains approximately 24 hours of audio of English speech and corresponding text transcriptions from one female speaker. 

\subsection{Model Configurations}

\subsubsection{Pre-training Configurations}
The Mixed-Phoneme BERT is made up of 8 feed-forward Transformer (FFT) blocks \cite{ren2019fastspeech} with a hidden size of 512 and 8 self-attention heads. Training of Mixed-Phoneme BERT was done on 24 Tesla V100-16GB GPUs, following the implementation and configuration suggested in \cite{liu2019roberta}. The model was trained with a batch size of 3072 sequences and a maximum length of 512 tokens for 220,000 steps, approximately 40 epochs. The BPE base dictionary was built using pre-training text data. Two different sizes of the BPE base dictionary, 3,000 and 30,000, were attempted in this study.

\subsubsection{TTS Finetuning Configurations}
After pre-training, the downstream TTS system based on the FastSpeech 2\cite{ren2020fastspeech} is warm-started on the text and mel-spectrogram pairs. FastSpeech 2 is an end-to-end non-autoregressive TTS approach, which consists of  three  components:  
a phoneme encoder network that encodes phoneme sequence as the text representation, a variance predictor that predicts the prosodic variance information, and a decoder that produces predicted mel-spectrograms. We replace the phoneme encoder in FastSpeech 2 with the Mixed-Phoneme BERT, and its parameters are initialized from the pre-trained model. The other components of FastSpeech 2 follow the original structure, and the parameters are initialized randomly.

The TTS front-end normalizes the sentences and converts them into phoneme sequences. The Parallel WaveGAN (PWG) \cite{yamamoto2020parallel} is used to generate speech waveform from the predicted mel-spectrograms. 

\subsubsection{Subjective Evaluation Configurations}
All the subjective tests are evaluated via our internal crowdsourced listening test platform. Total $60$ text sentences are selected for the subjective test, with at least $15$ native judges for the five-score MOS test and $12$  judges for Comparison MOS (CMOS)\cite{loizou2011speech}  for each test case. The CMOS gap over 0.05 suggests a significant gain. We also provide some demos of synthesized audios for reference.
 \footnote{https://speechresearch.github.io/mpbert}.

\vspace{-0.5em}

\section{Results and Analyses}

\subsection{TTS Performance Evaluation}
In \autoref{eval_table}, different TTS systems are compared to show the performance of pre-trained Mixed-Phoneme BERT. We compare the MOS and CMOS scores of audio samples generated by the following systems: i) \textit{GT}: ground-truth original speech; ii) \textit{GT (Mel+PWG)}: copy synthesis by Parallel WaveGAN vocoder with ground-truth mel-spectrograms; iii) \textit{FS2}: FastSpeech 2 with the default settings in \cite{ren2020fastspeech}; iv) \textit{FS2 w/o pre-trained MP BERT}: FastSpeech 2 with Mixed-Phoneme BERT as the phoneme encoder, which is not pre-trained; v) \textit{FS2 w/ pre-trained MP BERT}: FastSpeech 2 with pre-trained Mixed-Phoneme BERT as the phoneme encoder. The Mixed-Phoneme BERT is pre-trained with the whole word masking strategy. The base dictionary size for the BPE merge rules learning is 30,000. 

\begin{table}[htbp]
        \captionsetup{font=scriptsize}

        \caption{TTS Performance Evaluation. The MOS with 95\% confidence intervals.}
    \centering
    \scalebox{0.75}{
    \begin{tabular}{m{3.4cm}m{1.6cm}m{1.2cm}m{3cm}}
         \toprule 
         Method & MOS & CMOS (vs. \textit{FS2}) & CMOS (vs. \textit{FS2 w/o pre-trained MP BERT})  \\
           \midrule
         \textit{GT} & $4.17\pm0.07$   & -  & - \\
         \textit{GT(Mel+PWG)} & $4.02\pm0.08$   & -  & -  \\
         \midrule
         \textit{FS2}  & $3.75\pm0.06$   & 0.00  & - \\
         \textit{FS2 w/o pre-trained MP BERT}  & $3.90\pm0.08$   & 0.18  & 0.00  \\
         \midrule
         \textit{FS2 w/ pre-trained MP BERT}  & $\mathbf{4.04\pm0.07}$  & 0.30  & 0.18 \\
         \bottomrule
    \end{tabular}
    }
    \label{eval_table}
\end{table}

 It can be seen that using pre-trained Mixed-Phoneme BERT as the phoneme encoder (\textit{FS2 w/ pre-trained MP BERT}) can significantly improve the voice quality for Fastspech 2 compared with the systems \textit{FS2 w/o pre-trained MP BERT} and \textit{FS2}.  From the comments from the raters, we find the improvement on \textit{FS2 w/ pre-trained MP BERT} mainly results from the more expressive and appropriate prosody for the generated speech. 
It is also shown that system \textit{FS2 w/o pre-trained MP BERT} demonstrates better TTS performance than the \textit{FS2} system. The improvement is believed to result from the extra sup-phoneme information in the system inputs.

\vspace{-1em}

\subsubsection{Compared with the PnG BERT}
We evaluate the voice quality and inference latency of Mixed-Phoneme BERT compared with the recent TTS pre-trained model, PnG BERT\cite{jia21_interspeech}, which has a similar number of model parameters and training steps to Mixed-Phoneme BERT. We show the CMOS results and inference speedup for mel-spectrogram generation in \autoref{comp_pngbert}. Compared with PnG BERT, our model speeds up mel-spectrogram inference generation by $3\times$. The CMOS gain of the proposed model over PnG BERT is 0.03, showing that our proposed model performs on par or slightly better with PnG BERT. We also found that the training time of PnG BERT is about three times longer than the Mixed-Phoneme BERT~\footnote{Although PnG BERT and our Mixed-Phoneme BERT are trained with the same steps, PnG BERT concatenates character and phoneme sequence together as a longer one, which results in longer one-step forward time.}. 

\begin{table}[htbp]
        \captionsetup{font=scriptsize}

        \caption{RTF denotes the real-time factor, which is the time (in seconds) required for the system to synthesize a one-second waveform. The training and inference latency test is conducted on a server with 1 NVIDIA V100 GPU. }
    \centering
    \scalebox{0.8}{
    \begin{tabular}{m{1.5cm}m{2.9cm}m{2.5cm}m{1.1cm}}
         \toprule 
          Method & Inference Speed (RTF) & Inference Speedup  & CMOS \\
         \midrule
         \textit{PnG BERT}   & $3.99 \times 10^{-2}$  & -  & 0.00 \\
         \textit{MP BERT}   & $\mathbf{1.32 \times 10^{-2}}$  & $3\times$ & 0.03 \\
         \bottomrule
    \end{tabular}
    }
    \label{comp_pngbert}
\end{table}

\vspace{-1em}

\subsection{Ablation Studies}
\subsubsection{The Effectiveness of Mixed Phoneme and Sup-Phoneme Representations}
The effectiveness of using mixed representations in Mixed-Phoneme BERT is evaluated with the masked tokens prediction on the test set and TTS performance, shown in \autoref{effect_mpsr}. The pre-trained models are applied to predict the masked input tokens, and the masking ratio for input tokens is 15\%, which is the same as the training time. Two pre-trained models are involved for evaluation:  i) \textit{Phoneme BERT}: The Phoneme BERT only uses phoneme representations as model input; ii) \textit{MP BERT}: Mixed-Phoneme BERT using proposed mixed representations as model input.

\begin{table}[htbp]
        \captionsetup{font=scriptsize}

        \caption{Evaluations for the Mixed-Phoneme BERT and Phoneme BERT. Masked tokens prediction accuracy (MLM$_{Acc}$), measured at phoneme and sup-phoneme level.}
    \centering
    \scalebox{0.8}{
    \begin{tabular}{m{2.1cm}m{2.4cm}m{3.2cm}m{0.6cm}}
         \toprule 
          Method & MLM$_{Acc}$(phoneme) & MLM$_{Acc}$(sup-phoneme)  & CMOS \\
         \midrule
         \textit{Phoneme BERT}   & 45.40\%  & -  & 0.00\\
         \textit{MP BERT}   & \textbf{70.55\%}  & \textbf{70.43\%} & 0.13 \\
         \bottomrule
    \end{tabular}
    }
    \label{effect_mpsr}
\end{table}

It is shown that proposed mixed representations can help the pre-trained model significantly improve the masked tokens prediction accuracy. The subjective evaluation results show that using pre-trained Mixed-Phoneme BERT with mixed representations as input is more effective for the TTS fine-tuning than using Phoneme BERT only with phoneme representations. The objective and subjective improvements should result from that the sup-phoneme enhance the representation capacity of the Mixed-Phoneme BERT.

\subsubsection{The Effectiveness of Fine-Tuning Strategy}
In this section, three pre-trained Mixed-Phoneme BERT with different fine-tuning strategies are also investigated: i) \textit{MP BERT}: 15\% of input tokens are masked for the MLM prediction evaluation. The mixed phoneme and sup-phoneme representations are used for TTS fine-tuning; ii) \textit{MP BERT w/o sup-phoneme}: For MLM prediction evaluation, the masking ratio for phoneme tokens is 15\%, while all sup-phoneme tokens are masked. Then, the masked phoneme tokens and corresponding sup-phoneme tokens are predicted from the remained unmasked phoneme tokens in input. During the TTS fine-tuning stage, only the phoneme representations are used;  iii) \textit{MP BERT w/o phoneme}: For MLM prediction evaluation, the masking ratio for the sup-phoneme tokens is 15\%, while the entire phoneme sequence is masked. Furthermore, we only use sup-phoneme representations as inputs at TTS fine-tuning stage.

\begin{table}[htbp]
        \captionsetup{font=scriptsize}

        \caption{Evaluations for the Mixed-Phoneme BERT with three different fine-tuning strategies. Masked tokens prediction accuracy (MLM$_{Acc}$), measured at phoneme and sup-phoneme level.}
    \centering
    \scalebox{0.8}{
    \begin{tabular}{m{2.0cm}m{2.4cm}m{3.2cm}m{0.9cm}}
         \toprule 
          Method & MLM$_{Acc}$(phoneme) & MLM$_{Acc}$(sup-phoneme) & CMOS \\
         \midrule
         \textit{MP BERT}   & \textbf{70.55\%}  & \textbf{70.43\%} & 0 \\
         \midrule
         \textit{- phoneme}   & 59.78\%  & 54.44\%  & -0.10 \\
         \textit{- sup-phoneme}   & 29.24\%   & 25.65\% & -0.11 \\
         \bottomrule
    \end{tabular}
    }
    \label{effect_fts}
\end{table}

The results for MLM prediction evaluation and TTS performance are shown in \autoref{effect_fts}. It shows that both masked token prediction accuracy and TTS performance drop when the phoneme or sup-phoneme representations are not used. Specifically, we find the performance degradation is severer when the Mixed-Phoneme BERT only uses phoneme rather than sup-phoneme representations, suggesting the effectiveness of sup-phoneme representations.

\vspace{-0.5em}

\subsubsection{Analysis on Masking Strategy}
We also study the effect of the masking strategy on the performance of the Mixed-Phoneme BERT, as shown in \autoref{masking_effect}. The subjective evaluation results show that the whole word masking strategy increases TTS performance. The work \cite{jia21_interspeech} also shows a similar discovery.
We consider that when the representation capacity of the model input is not changed, increasing the difficulty of the MLM prediction task to some extent might improve the performance of the downstream TTS task.

\begin{table}[htbp]
        \captionsetup{font=scriptsize}

        \caption{Evaluations for the Mixed-Phoneme BERT with different input masking strategies. Masked tokens prediction accuracy (MLM$_{Acc}$), measured at phoneme and sup-phoneme level.  WWM means whole word masking.}
    \centering
    \scalebox{0.8}{
    \begin{tabular}{m{2.0cm}m{2.4cm}m{3.2cm}m{0.9cm}}
         \toprule 
          Strategy & MLM$_{Acc}$(phoneme) & MLM$_{Acc}$(sup-phoneme)  & CMOS \\
         \midrule
         \textit{w/ WWM}   & 70.55\%  & 70.43\%   & 0.00 \\
         \midrule
         \textit{w/o WWM}   & \textbf{74.69\%}   & - & -0.08 \\
         \bottomrule
    \end{tabular}
    }
    \label{masking_effect}
\end{table}

\vspace{-1em}
\section{Conclusion}
This paper proposed Mixed-Phoneme BERT, a novel variant of the BERT model to handle TTS tasks. Specifically, we introduce mixed phoneme and sup-phoneme representations as to the input of the Mixed-Phoneme BERT. The mixed representations can enhance the representation capacity of the model while avoiding the inconsistency between pre-training and fine-tuning. The subjective evaluations show that the pre-trained Mixed-Phoneme BERT can improve the performance of FastSpeech 2 significantly and enjoy fast inference speed.

% \section{Acknowledge}
% This research is partially supported by a Knowledge Transfer Project Fund (Ref: KPF20QEP26) from the Chinese University of Hong Kong.

\vfill\pagebreak

\bibliographystyle{IEEEbib}
\bibliography{strings,refs}

\begin{thebibliography}{10}
\providecommand{\url}[1]{#1}
\csname url@samestyle\endcsname
\providecommand{\newblock}{\relax}
\providecommand{\bibinfo}[2]{#2}
\providecommand{\BIBentrySTDinterwordspacing}{\spaceskip=0pt\relax}
\providecommand{\BIBentryALTinterwordstretchfactor}{4}
\providecommand{\BIBentryALTinterwordspacing}{\spaceskip=\fontdimen2\font plus
\BIBentryALTinterwordstretchfactor\fontdimen3\font minus
  \fontdimen4\font\relax}
\providecommand{\BIBforeignlanguage}[2]{{%
\expandafter\ifx\csname l@#1\endcsname\relax
\typeout{** WARNING: IEEEtran.bst: No hyphenation pattern has been}%
\typeout{** loaded for the language `#1'. Using the pattern for}%
\typeout{** the default language instead.}%
\else
\language=\csname l@#1\endcsname
\fi
#2}}
\providecommand{\BIBdecl}{\relax}
\BIBdecl

\bibitem{tan2021survey}
X.~Tan, T.~Qin, F.~Soong, and T.-Y. Liu, ``A survey on neural speech
  synthesis,'' \emph{arXiv preprint arXiv:2106.15561}, 2021.

\bibitem{tan2022naturalspeech}
X.~Tan, J.~Chen, H.~Liu, J.~Cong, C.~Zhang, Y.~Liu, X.~Wang, Y.~Leng, Y.~Yi,
  L.~He \emph{et~al.}, ``Naturalspeech: End-to-end text to speech synthesis
  with human-level quality,'' \emph{arXiv preprint arXiv:2205.04421}, 2022.

\bibitem{wang2017tacotron}
Y.~Wang, R.~Skerry-Ryan, D.~Stanton, Y.~Wu, R.~J. Weiss, N.~Jaitly, Z.~Yang,
  Y.~Xiao, Z.~Chen, S.~Bengio \emph{et~al.}, ``Tacotron: Towards end-to-end
  speech synthesis,'' in \emph{Proc. Interspeech}, Aug. 2017, pp. 4006--4010.

\bibitem{shen2018natural}
J.~{Shen}, R.~{Pang}, R.~J. {Weiss}, M.~{Schuster}, N.~{Jaitly}, Z.~{Yang},
  Z.~{Chen}, Y.~{Zhang}, Y.~{Wang}, R.~{Skerrv-Ryan}, R.~A. {Saurous},
  Y.~{Agiomvrgiannakis}, and Y.~{Wu}, ``Natural tts synthesis by conditioning
  wavenet on mel spectrogram predictions,'' in \emph{Proc. ICASSP}, 2018, pp.
  4779--4783.

\bibitem{ren2019fastspeech}
Y.~Ren, Y.~Ruan, X.~Tan, T.~Qin, S.~Zhao, Z.~Zhao, and T.-Y. Liu, ``Fastspeech:
  Fast, robust and controllable text to speech,'' in \emph{Proc. NIPS}, 2019,
  pp. 3171--3180.

\bibitem{ren2020fastspeech}
Y.~Ren, C.~Hu, X.~Tan, T.~Qin, S.~Zhao, Z.~Zhao, and T.-Y. Liu, ``Fastspeech 2:
  Fast and high-quality end-to-end text to speech,'' in \emph{Proc. ICLR},
  2021.

\bibitem{zhang2021study}
G.~Zhang, Y.~Leng, D.~Tan, Y.~Qin, K.~Song, X.~Tan, S.~Zhao, and T.~Lee, ``A
  study on the efficacy of model pre-training in developing neural
  text-to-speech system,'' \emph{arXiv preprint arXiv:2110.03857}, 2021.

\bibitem{elias21_interspeech}
I.~Elias, H.~Zen, J.~Shen, Y.~Zhang, Y.~Jia, R.~Skerry-Ryan, and Y.~Wu,
  ``{Parallel Tacotron 2: A Non-Autoregressive Neural TTS Model with
  Differentiable Duration Modeling},'' in \emph{Proc. Interspeech}, 2021, pp.
  141--145.

\bibitem{zhang21u_interspeech}
G.~Zhang, Y.~Qin, D.~Tan, and T.~Lee, ``{Applying the Information Bottleneck
  Principle to Prosodic Representation Learning},'' in \emph{Proc.
  Interspeech}, 2021, pp. 3156--3160.

\bibitem{kenter2020improving}
T.~Kenter, M.~Sharma, and R.~Clark, ``Improving the prosody of rnn-based
  english text-to-speech synthesis by incorporating a bert model,'' in
  \emph{Proc. Interspeech}, 2020, pp. 4412--4416.

\bibitem{taylor2009text}
P.~A. Taylor, \emph{Text-to-speech synthesis}.\hskip 1em plus 0.5em minus
  0.4em\relax Cambridge University Press, 2009.

\bibitem{zen2015unidirectional}
H.~Zen and H.~Sak, ``Unidirectional long short-term memory recurrent neural
  network with recurrent output layer for low-latency speech synthesis,'' in
  \emph{Proc. ICASSP}.\hskip 1em plus 0.5em minus 0.4em\relax IEEE, 2015, pp.
  4470--4474.

\bibitem{wu2016merlin}
Z.~Wu, O.~Watts, and S.~King, ``Merlin: An open source neural network speech
  synthesis system.'' in \emph{SSW}, 2016, pp. 202--207.

\bibitem{zhang2020learning}
G.~Zhang, Y.~Qin, and T.~Lee, ``Learning syllable-level discrete prosodic
  representation for expressive speech generation,'' in \emph{Proc.
  Interspeech}, 2020, pp. 3426--3430.

\bibitem{wang2015word}
P.~Wang, Y.~Qian, F.~K. Soong, L.~He, and H.~Zhao, ``Word embedding for
  recurrent neural network based tts synthesis,'' in \emph{Proc. ICASSP}.\hskip
  1em plus 0.5em minus 0.4em\relax IEEE, 2015, pp. 4879--4883.

\bibitem{chung2019semi}
Y.-A. Chung, Y.~Wang, W.-N. Hsu, Y.~Zhang, and R.~Skerry-Ryan,
  ``Semi-supervised training for improving data efficiency in end-to-end speech
  synthesis,'' in \emph{Proc. ICASSP}.\hskip 1em plus 0.5em minus 0.4em\relax
  IEEE, 2019, pp. 6940--6944.

\bibitem{hayashi2019pre}
T.~Hayashi, S.~Watanabe, T.~Toda, K.~Takeda, S.~Toshniwal, and K.~Livescu,
  ``Pre-trained text embeddings for enhanced text-to-speech synthesis.'' in
  \emph{Proc. Interspeech}, 2019, pp. 4430--4434.

\bibitem{xu2021improving}
G.~Xu, W.~Song, Z.~Zhang, C.~Zhang, X.~He, and B.~Zhou, ``Improving prosody
  modelling with cross-utterance bert embeddings for end-to-end speech
  synthesis,'' in \emph{Proc. ICASSP}.\hskip 1em plus 0.5em minus 0.4em\relax
  IEEE, 2021, pp. 6079--6083.

\bibitem{devlin2019bert}
J.~Devlin, M.-W. Chang, K.~Lee, and K.~Toutanova, ``Bert: Pre-training of deep
  bidirectional transformers for language understanding,'' in \emph{Proc.
  NAACL}, 2019, pp. 4171--4186.

\bibitem{xiao2020improving}
Y.~Xiao, L.~He, H.~Ming, and F.~K. Soong, ``Improving prosody with linguistic
  and bert derived features in multi-speaker based mandarin chinese neural
  tts,'' in \emph{Proc. ICASSP}.\hskip 1em plus 0.5em minus 0.4em\relax IEEE,
  2020, pp. 6704--6708.

\bibitem{el2020characterbert}
H.~El~Boukkouri, O.~Ferret, T.~Lavergne, H.~Noji, P.~Zweigenbaum, and
  J.~Tsujii, ``Characterbert: Reconciling elmo and bert for word-level
  open-vocabulary representations from characters,'' in \emph{International
  Conference on Computational Linguistics}, 2020, pp. 6903--6915.

\bibitem{liu2019roberta}
Y.~Liu, M.~Ott, N.~Goyal, J.~Du, M.~Joshi, D.~Chen, O.~Levy, M.~Lewis,
  L.~Zettlemoyer, and V.~Stoyanov, ``Roberta: A robustly optimized bert
  pretraining approach,'' \emph{arXiv preprint arXiv:1907.11692}, 2019.

\bibitem{kastner2019representation}
K.~Kastner, J.~F. Santos, Y.~Bengio, and A.~Courville, ``Representation mixing
  for tts synthesis,'' in \emph{Proc. ICASSP}.\hskip 1em plus 0.5em minus
  0.4em\relax IEEE, 2019, pp. 5906--5910.

\bibitem{jia21_interspeech}
Y.~Jia, H.~Zen, J.~Shen, Y.~Zhang, and Y.~Wu, ``{PnG BERT: Augmented BERT on
  Phonemes and Graphemes for Neural TTS},'' in \emph{Proc. Interspeech 2021},
  2021, pp. 151--155.

\bibitem{ding2019call}
S.~Ding, A.~Renduchintala, and K.~Duh, ``A call for prudent choice of subword
  merge operations in neural machine translation,'' in \emph{Proceedings of
  Machine Translation Summit XVII: Research Track}, 2019, pp. 204--213.

\bibitem{sennrich2016neural}
R.~Sennrich, B.~Haddow, and A.~Birch, ``Neural machine translation of rare
  words with subword units,'' in \emph{Proc. ACL}, 2016, pp. 1715--1725.

\bibitem{wolf2020transformers}
T.~Wolf, J.~Chaumond, L.~Debut, V.~Sanh, C.~Delangue, A.~Moi, P.~Cistac,
  M.~Funtowicz, J.~Davison, S.~Shleifer \emph{et~al.}, ``Transformers:
  State-of-the-art natural language processing,'' in \emph{Proc. EMNLP}, 2020,
  pp. 38--45.

\bibitem{chinese-bert-wwm}
Y.~Cui, W.~Che, T.~Liu, B.~Qin, Z.~Yang, S.~Wang, and G.~Hu, ``Pre-training
  with whole word masking for chinese bert,'' \emph{arXiv preprint
  arXiv:1906.08101}, 2019.

\bibitem{yamamoto2020parallel}
R.~Yamamoto, E.~Song, and J.-M. Kim, ``Parallel wavegan: A fast waveform
  generation model based on generative adversarial networks with
  multi-resolution spectrogram,'' in \emph{Proc. ICASSP}.\hskip 1em plus 0.5em
  minus 0.4em\relax IEEE, 2020, pp. 6199--6203.

\bibitem{loizou2011speech}
P.~C. Loizou, ``Speech quality assessment,'' in \emph{Multimedia analysis,
  processing and communications}.\hskip 1em plus 0.5em minus 0.4em\relax
  Springer, 2011, pp. 623--654.

\end{thebibliography}

\end{document}